\documentclass[%
 reprint,
 amsmath,amssymb,
 aps,
 pre,
]{revtex4-2}
\usepackage{soul}
\usepackage{graphicx}% Include figure files
\usepackage{dcolumn}% Align table columns on decimal point
\usepackage{bm}% bold math
\usepackage{xcolor}% 
\usepackage{enumitem}

\newcommand{\grad}{\nabla}

\newcommand{\laplacian}{\nabla^2}

\newcommand{\eq}[1]{eq.~(\ref{#1})}

\begin{document}

\preprint{APS/123-QED}

\title{The Unsteady Taylor--Vortex Dynamo is Fast}

\author{Liam O'Connor$^{1,2}$, Daniel Lecoanet$^{1,2}$, Geoffrey M. Vasil$^{3}$, Kyle C. Augustson$^{4}$, Florentin Daniel$^{2}$, Evan H. Anders$^{5}$, Keaton J. Burns$^{6,7}$, Jeffrey S. Oishi$^{8}$, Benjamin P. Brown$^{9}$}

\address{$^1$Department of Engineering Sciences and Applied Mathematics, Northwestern University, Evanston, IL, 60208 USA}
\address{$^2$Center for Interdisciplinary Exploration and Research in Astrophysics, Northwestern University, Evanston, IL, 60201 USA}
\address{$^3$School of Mathematics, University of Edinburgh, EH9 3FD, UK}
\address{$^4$Southwest Research Institute, 1301 Walnut Street, Suite 400, Boulder, CO, 80302 USA}
\address{$^5$Kavli Institute for Theoretical Physics, University of California, Santa Barbara, Santa Barbara, CA, 93106, USA}
\address{$^6$Department of Mathematics, Massachusetts Institute of Technology, Cambridge, MA, 02142 USA}
\address{$^7$Center for Computational Astrophysics, Flatiron Institute, New York, NY, 10010 USA}
\address{$^8$Department of Mechanical Engineering, University of New Hampshire, Kingsbury Hall, Durham, NH, 03824 USA}
\address{$^9$Department of Astrophysical and Planetary Sciences, University of Colorado Boulder, Boulder, CO, 80309 USA}

\begin{abstract}
Astrophysical and geophysical fluids commonly generate organized magnetic fields, despite having enormous magnetic Reynolds numbers $\rm{Rm}$ and abundant small-scale turbulence. 
Flow-induced dynamo action produces these fields, with the ``kinematic dynamo problem'' devoted to determining the rate at which a flow exponentially amplifies weak magnetic fields. 
However, previous studies on high-Rm kinematic dynamos have generated flows via imposed volumetric forcing or oscillatory boundary conditions. 
In this letter, we investigate a system with  three important attributes: realistic flow conditions, fast dynamo action (operational for $\rm{Rm}\to\infty$), and a subharmonic spatio-temporal structure. 
We show that unsteady Taylor--vortex flow, a  regime observed in laboratory experiments, gives rise to fast dynamos with time and length scales twice those of the flow at high $\rm{Rm}$. 
By numerically integrating a Floquet system driven by periodic oscillations of Taylor vortices, we solve the kinematic dynamo problem up to $\rm{Rm} = 3.2\cdot 10^6$, calculating the dynamo's growth rate as a function of Rm and streamwise wavenumber. 
We find the onset of instability and compute Finite-Time Lyapunov Exponents, which identify the regions of Lagrangian chaos required for fast dynamo action. 
To our knowledge, unsteady Taylor--vortex flow produces the most physically motivated fast dynamo to date. 
\end{abstract}
\maketitle

The inductive motions of highly conducting fluids in stars and planets generate magnetic fields by converting a flow's kinetic energy into magnetic energy.
These fields originate from instabilities related to a preexisting ``fossil'' field or from the exponential amplification of weak seed fields.
The latter case can be examined by solving the kinematic dynamo equations where the field's amplitude is assumed to be small.

Kinematic dynamos are classified as ``fast'' provided they amplify magnetic fields at infinite magnetic Reynolds number ($\rm{Rm}\to\infty$). 
Fast dynamos are particularly relevant to astrophysical systems where $\rm{Rm}$ is generally very large.
Unlike slow dynamos, which are characterized by their long diffusive time scales, fast dynamos are characterized by their shorter advective time scales.
Fast dynamo action is necessary to explain the rapid magnetic field regeneration observed in stellar activity cycles, including the 11-year solar cycle.

Early studies of fast kinematic dynamos have largely focused on contrived maps \cite{bayly_fast_1987,bayly_construction_1988, 10.1063/1.866956, childress_fast_1992} and idealized flows \cite{cattaneo_fluctuations_1995,HUGHES1996167,hollerbach_numerical_1995, otani_fast_1993}.
Periodic ABC-type flows in particular have received abundant analytical and numerical consideration \cite{zheligovsky_numerical_1993,gilbert_fast_1991, tanner_fast-dynamo_2003,galloway_numerical_1992,bouya_toward_2015}.
Later efforts focused on fast dynamos driven by flows with more physical significance.
\cite{smith_vortex_2004} provides evidence for a fast vortex dynamo sustained by physically motivated volumetric forcing.
\cite{seshasayanan_turbulent_2016} uses a similar approach in the turbulent regime, demonstrating fast dynamo action in another volumetrically driven flow.
 
Especially relevant to the present study, a fast dynamo has also been obtained using a realistic flow generated by solving the Navier--Stokes equations with time-periodic boundary conditions \cite{khalzov_fast_2013}.
Despite their use of physical flows, these dynamos rely on volumetric forcing and oscillatory boundary conditions, making them difficult to achieve experimentally.

In this work, we present evidence for a fast dynamo in  a flow that has previously been realized and characterized experimentally.
We describe how Taylor vortices, which have been observed in laboratory experiments of rotating shear flows, amplify magnetic seeds by periodically stretching field lines in regions of Lagrangian chaos.
Although dynamos driven by Taylor vortices have been investigated \cite{marcotte_dynamo_2016,gissinger_taylor-vortex_2014,willis_taylor-couette_2002,laguerre_cyclic_2008,laure_generation_2001}, these efforts did not classify the system as fast or slow.
We simulate a kinematic dynamo up to $\rm{Rm}=3.2\cdot 10^6$ while computing the magnetic field's growth rate and Floquet mode as a function of $\rm{Rm}$ and streamwise wavenumber.
The resulting dynamos exhibit spatiotemporal subharmonic structure, with magnetic fields that evolve on twice the spatial and temporal scales of the flow at high $\rm{Rm}$.
We also compute Finite-Time Lyapunov Exponents (FTLEs) to identify regions of Lagrangian chaos, a necessary condition for fast dynamos.
Finally, we demonstrate that the dynamo's effective wavenumber $k_{\rm{eff}}\sim \rm{Rm}^{1/2}$ in agreement with \cite{khalzov_fast_2013}.
These results have implications for the experimental design of dynamos in laboratories as well as the spontaneous amplification of large-scale magnetic fields in stars and planets.

\begin{figure*}
    \centering
    \includegraphics[width=6.8in]{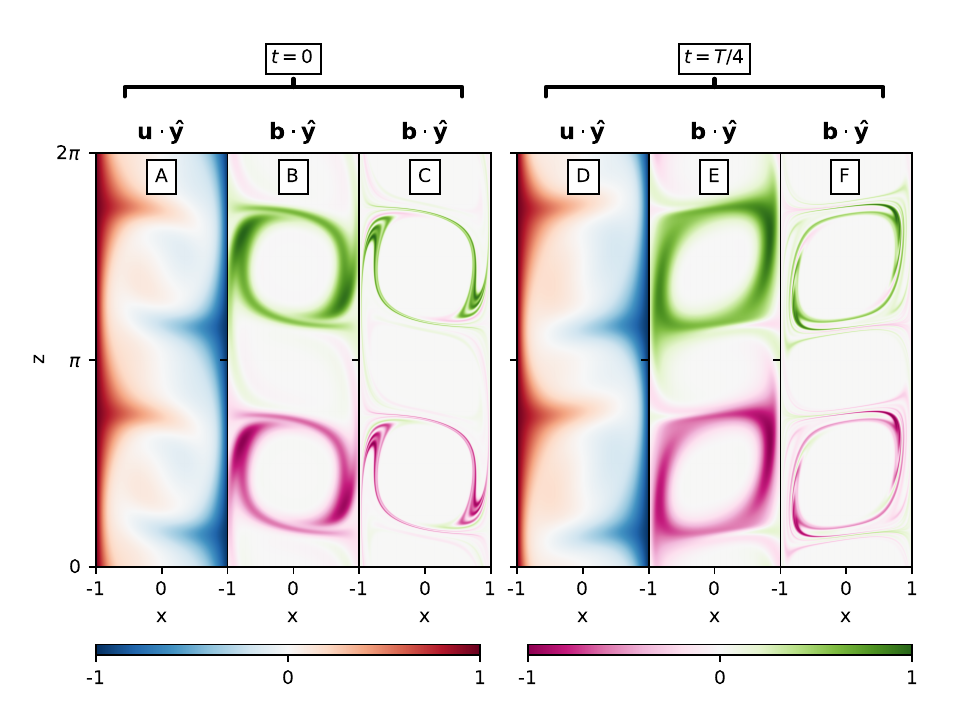}
    \caption{Snapshots of the dynamo cycle are shown at the beginning ($t=0$, panes A, B, and C) and after one quarter of a period ($t=T/4$, panes D, E, and F). 
    At both times, we plot the streamwise velocity (panes A and D) and streamwise magnetic field 
    at $\rm{Rm} = 1.5\cdot 10^3$ (panes B and E) and $\rm{Rm} = 1.5\cdot 10^5$ (panes C and F).
    Both dynamos were generated at streamwise wavenumber $k_y = 0.18$.
    We observe Taylor vortices in the streamwise velocity field, given by two pairs of counter-rotating cells which transport momentum between thin boundary layers. 
    As the plumes of these cells oscillate in $z$, the magnetic field is amplified in ring-like structures in alternating regions between plumes, such that the rings co-rotate.
    At $t=0$, we observe two rings rotating clockwise. 
    At $t=T/4$, there is a transition toward counterclockwise rotating rings which form in regions where the magnetic field was initially weak.
    The $\rm{Rm}=1.5\cdot 10^3$ dynamo (panes B and E) has a more diffuse structure than $\rm{Rm}=1.5\cdot 10^5$ (panes C and F).
    Both dynamos are dominated by the $k_z = 1$ mode, whereas the velocity field is primarily $k_z = 2$.
    }
    \label{big}
\end{figure*}

We model the dynamo using a rotating shear flow in a 2.5D cartesian box. 
The velocity $\mathbf{u}(x, z, t)$ has 2D spatial dependence with components in three directions: $\mathbf{\hat{x}}$ (wall-normal), $\mathbf{\hat{y}}$ (streamwise), and $\mathbf{\hat{z}}$ (spanwise).
Our problem domain has $-1 < x < 1$ and $0 < z < 2\pi$ where $\mathbf{u}|_{x=\pm 1} =\mp \mathbf{\hat{y}}$ with periodicity in $z$. 

Although the magnetic field $\mathbf{b}$ is fully 3D, we only consider sinusoidal streamwise dependence such that $\mathbf{b}$ can be expressed as $\mathbf{b}(x, y, z, t) = \Re\left[\mathbf{B}(x,z,t)\exp\left(i k_y y \right) \right]$ where $\mathbf{B} \in \mathbb{C}^3$ and $k_y$ denotes the streamwise wavenumber.
This implies $\mathbf{b}$ has no net streamwise flux.
We enforce the same constraint in $z$ by requiring $\int_{0}^{2\pi} \mathbf{b}\cdot \mathbf{\hat{z}} \,\rm{d}z = 0$.
In $x$, there is no net flux due to perfectly conducting boundary conditions ($\mathbf{b}\cdot\mathbf{\hat{x}}|_{x=\pm 1} = 0$).

The rotating kinematic equations are given by
\begin{align}
    \grad\cdot\mathbf{u} = \grad\cdot\mathbf{b} &= 0\label{divEQ}\\
    \frac{\partial\mathbf{u}}{\partial t} + (\mathbf{u}\cdot\grad)\mathbf{u} + \grad p + \frac{\mathbf{\hat{z}}\times\mathbf{u}}{\rm{Ro}} &= \frac{1}{\rm{Re}}\laplacian \mathbf{u} \label{nsEQ}\\
    \frac{\partial\mathbf{b}}{\partial t} - \grad\times(\mathbf{u}\times\mathbf{b}) &= \frac{1}{\rm{Rm}}\laplacian\mathbf{b}\label{indEQ}
\end{align}
where Ro, Re, and Rm respectively denote the Rossby number, Reynolds number, and magnetic Reynolds number.
Length scales are nondimensionalized by the half-thickness of the domain in the wall-normal direction ($x$), denoted $L_x$.
The time coordinate $t$ is nondimensionalized by the advective turnover time such that Ro, Re, and Rm respectively scale with the rotation, viscous, and ohmic times.

To solve \eq{indEQ}, we evolve the vector potential $\mathbf{A}$, which defines the magnetic field as $\mathbf{b} = \grad\times\mathbf{A}$.
We enforce the Coulomb gauge, given by $\grad\cdot\mathbf{A}=0$. 
We solve this system numerically using the \texttt{Dedalus} pseudospectral solver \cite{Burns2020}, where the $x$-dependence ($z$-dependence) of each variable is represented using Chebyshev (Fourier) modes.
At low Rm, we use spatial resolutions of 64 modes in $x$ and 256 modes in $z$.
We uniformly timestep with $\Delta t=5\cdot 10^{-3}$ using a second-order Runge-Kutta method.
We increase the spatial and temporal resolutions progressively with increasing Rm, such that the highest-Rm case uses 1024 (4096) modes in $x$ ($z$) with $\Delta t = 2.5\cdot 10^{-4}$.
We conduct resolution tests for all cases with $\mathrm{Rm}>10^6$ and find that increasing the spatial and temporal resolutions by a factor of 1.5 produces no appreciable change in the magnetic growth rate.

We consider magnetic perturbations with streamwise wavenumbers $k_y \in \{0.18, \, 0.29, \, 1.0\}$. 
In cylindrical Taylor--Couette experiments, dynamo action is typically associated with the azimuthal mode $m=1$ \cite{marcotte_dynamo_2016,gissinger_taylor-vortex_2014,willis_taylor-couette_2002,laguerre_cyclic_2008,laure_generation_2001}. 
The values $k_y=0.18$ and $0.29$ correspond to the $m=1$ mode for cylinders of approximate radii $5L_x$ and $3L_x$, respectively; $k_y=1.0$ may represent a higher-$m$ mode.

Although cartesian rotating plane Couette flow (RPCF) is often portrayed as a simplified model for cylindrical Taylor--Couette flow, as these systems are equivalent in the narrow-gap limit \cite{nagata_taylorcouette_2023}, RPCF has also been the subject of dedicated numerical and experimental studies \cite{yang_bifurcation_2021,tsukahara_flow_2010}.
Laboratory experiments of RPCF carried out by \cite{tsukahara_flow_2010} displayed streamwise-symmetric Taylor vortices for $50 <\rm{Re} < 200$ and $\rm{Ro}>5$  (the minimum Rossby number of their study).

In our investigation, we decrease the Rossby number even further to $\rm{Ro}=3$ and fix $\rm{Re}=150$ while varying $\rm{Rm}$ (or, equivalently, the magnetic Prandtl number $\rm{Pm}\equiv \rm{Rm} / \rm{Re}$).
In contrast to \cite{tsukahara_flow_2010}, we find two stable states: a steady flow with a single vortex pair and an oscillatory flow with two vortex pairs.
Our 2.5D model is motivated by the fact that both states display streamwise symmetry in full 3D hydrodynamic simulations.

Although previous studies reported dynamo action in stationary Taylor--Couette flows \cite{willis_taylor-couette_2002, gissinger_taylor-vortex_2014}, we find that magnetic perturbations decay in the steady single-vortex-pair state. 
The magnetic field also decays when we freeze the oscillatory velocity field in time, a result consistent with \cite{tilgner_dynamo_2008}. 
The oscillatory flow studied here is likely a 2D meandering state similar to that described by \cite{tsukahara_flow_2010}, and we expect this mechanism to sustain dynamo action over a range of $\mathrm{Ro}$ and $\mathrm{Re}$ within this regime, as well as in other oscillatory Taylor--vortex flows.

Pane A of Figure~\ref{big} shows the streamwise velocity component of the oscillatory Taylor--vortex flow at $t=0$.
Adjacent counter-rotating vortices are separated by localized plumes of increased streamwise velocity which transport momentum in the wall-normal ($x$) direction.
These plumes oscillate periodically in $z$ such that their encompassed vortices expand and contract with a period of $T=23.33$.
At $t=0$, the vortex centered near $z=\pi$ has nearly expanded to its maximum size.
In pane D of Figure~\ref{big}, we illustrate the flow's evolution by plotting the streamwise velocity again at a later time $t=T/4$ at which point, the $z=\pi$ vortex begins to contract.

\begin{figure}[t]
    \centering
    \includegraphics[width=3.4in]{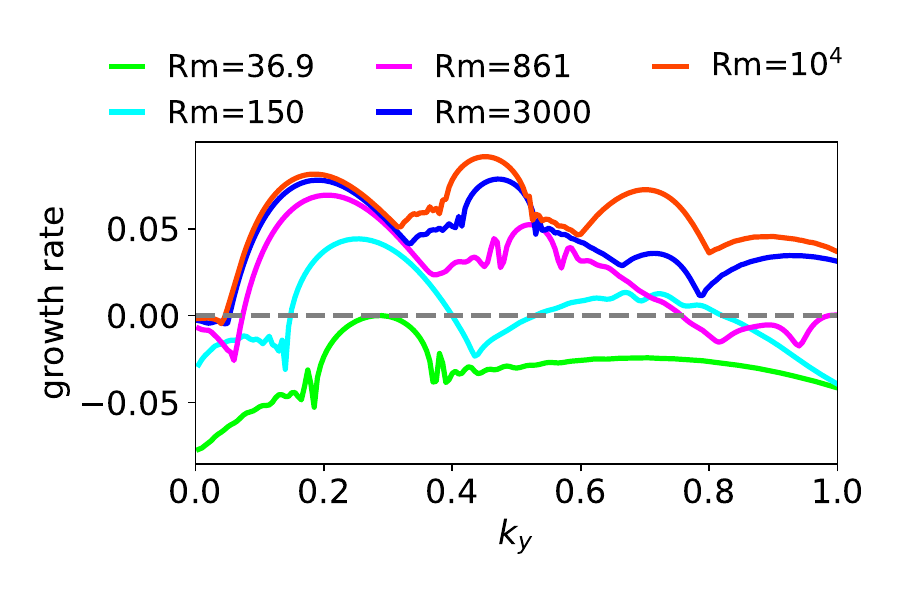}
    \caption{Growth rates vs. streamwise wavenumber ($k_y$) for various $\rm{Rm}$. 
    The onset of instability occurs at $\rm{Rm}=36.9$ and $k_y=0.29$ while the dominant $k_y=1.0$ mode becomes marginally stable when $\rm{Rm}=861$.
    As $\rm{Rm}$ increases, the maximizing $k_y$ decreases for individual extrema while higher $k_y$ modes become dominant.
    }
    \label{gamma_ky}
\end{figure}

While simulating this periodic flow, we solve a Floquet system via direct integration to isolate the dominant magnetic  Floquet mode.
We initialize the magnetic field using a random seed, then solve the linear system \eq{indEQ} until an exactly periodic structure emerges.
We evolve the system for at least 10 periods, observing exponential growth/decay of the magnetic field amplitude.
In Figure~\ref{big}, we plot snapshots of two dynamos' Floquet modes at the initial time ($t=0$, panes B and C) and after one quarter of a period ($t=T/4$, panes E and F).

We set $k_y=0.18$ in both cases.
The first dynamo (panes B and E), computed at $\rm{Rm}=1.5\cdot 10^3$, has a relatively diffuse magnetic structure while the second dynamo (panes C and F), computed at $\rm{Rm}=1.5\cdot 10^5$, has intricate small-scale features.
Despite this contrast, the regions of significant field coincide with approximately equal area in both cases.

The magnetic cycles of both dynamos synchronize spatially and temporally with vortex oscillations.
While vortices undergo expansion, ring-shaped structures of increased magnetic field form near the outer edges of the co-rotating vortices.
In panes B and C of Figure~\ref{big}, we observe one pair of magnetic rings rotating clockwise while their coinciding vortices approach maximum expansion.
These magnetic rings disappear as their host vortices begin to contract while another pair forms in the adjacent vortices.
Panes E and F of Figure~\ref{big} illustrates this transition, where the magnetic rings at $z=0$ and $z=\pi$ are disrupted while a new pair of rings begins to form at $z=\pi/2$ and $z=3\pi/2$.
The new pair of rings will rotate counterclockwise.
Notice how the magnetic field's sign reverses when comparing two rings at a single time. 
This finding agrees with previous investigations, where the magnetic field was observed to span twice the vertical lengthscale as the Taylor vortices \cite{gissinger_taylor-vortex_2014}.

\begin{figure}[t]
    \centering
    \includegraphics[width=3.4in]{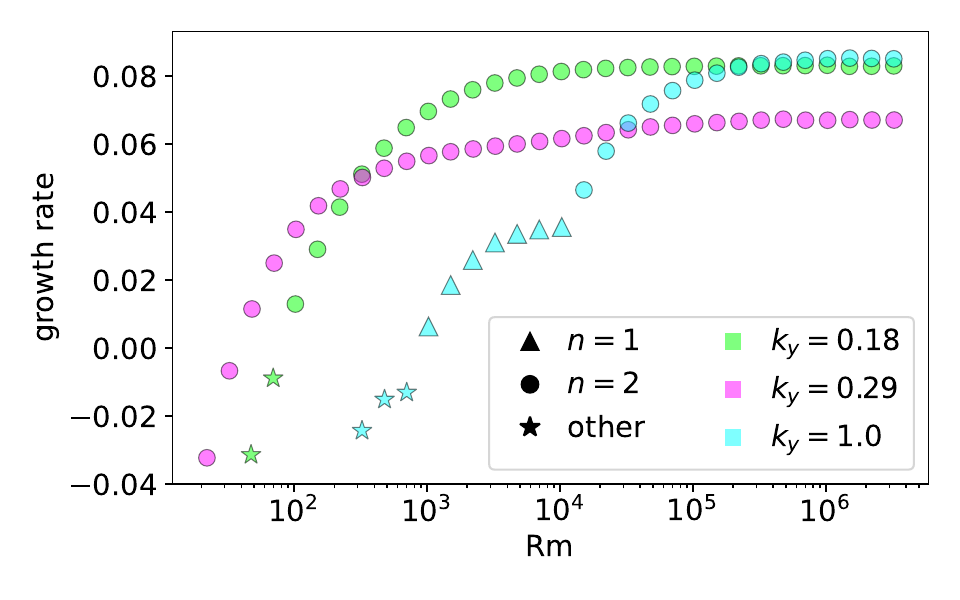}
    \caption{
    The magnetic growth rate is plotted as a function of Rm for streamwise wavenumbers $k_y=0.18, \, 0.29, \, 1.0$.
    For $k_y=1.0$, we observe cusp points where the growth rate's derivative with respect to $\rm{Rm}$ appears discontinuous.
    Evidence for a fast dynamo is observed at high $\rm{Rm}$, where the growth rate exhibits no appreciable dependence on $\rm{Rm}$.
    Triangle markers denote cases where the Floquet modes' period is equal to the flow's oscillation period $T$. 
    Circular points denote Floquet modes with period $2T$ and the cases marked by stars have even longer periods or exhibit quasiperiodicity.
    }
    \label{gamma_rm}
\end{figure}

In addition to the magnetic  Floquet modes, we compute the maximum growth rate of $\mathbf{b}$ using linear regression.
In Figure~\ref{gamma_ky}, we plot the magnetic field's growth rate as a function of $k_y$ for several $\rm{Rm}$.
The onset of instability is shown in green, where the maximum growth rate is zero for $\rm{Rm}=36.9$ and $k_y=0.29$. 
The critical magnetic Prandtl number $\rm{Pm}_{\rm{c}}=0.246$.
In general, the growth rate increases with $\rm{Rm}$ due to the diminishing influence of diffusion.
This trend is not always monotonic, as the growth rate for $\rm{Rm}=150$ exceeds that of $\rm{Rm}=861$ near $k_y=0.06$ and $k_y=0.8$.
The $k_y=1.0$ mode becomes marginally stable when $\rm{Rm}=861$.
As $\rm{Rm}$ increases, the maximizing $k_y$ of each extremum decreases.
For example, the critical mode's maximizer goes from $k_y=0.29$ at $\rm{Rm}=36.9$ to $k_y=0.21$ at $\rm{Rm}=861$.
As $\mathrm{Rm}$ increases, the maximum growth rate and corresponding $k_y$ of this extremum approach constant values, converging to $k_y = 0.18$ with a growth rate of 0.081 by $\mathrm{Rm} = 10^4$.
This convergence is consistent with fast dynamo action, providing a lower bound on the growth rate as $\mathrm{Rm} \to \infty$, even though other modes become more unstable at different $k_y$.
For instance, at $\mathrm{Rm} \gtrsim 3000$, a different extremum becomes dominant at $k_y = 0.47$.
We anticipate that this new dominant mode will likewise converge to a characteristic $k_y$ and growth rate as $\mathrm{Rm} \to \infty$.
Although the flow's oscillation period $T$ remains fixed, varying $k_y$ changes the streamwise advective turnover time $\tau_y = 2\pi/k_y$, thereby altering the nondimensional period $T/\tau_y$.
The resulting dependence of growth rate on $T/\tau_y$ indicates a preferred oscillation frequency, as also reported in \cite{dormy_time_2008,tilgner_dynamo_2008}.

\begin{figure}[t]
    \centering
    \includegraphics[width=3.4in]{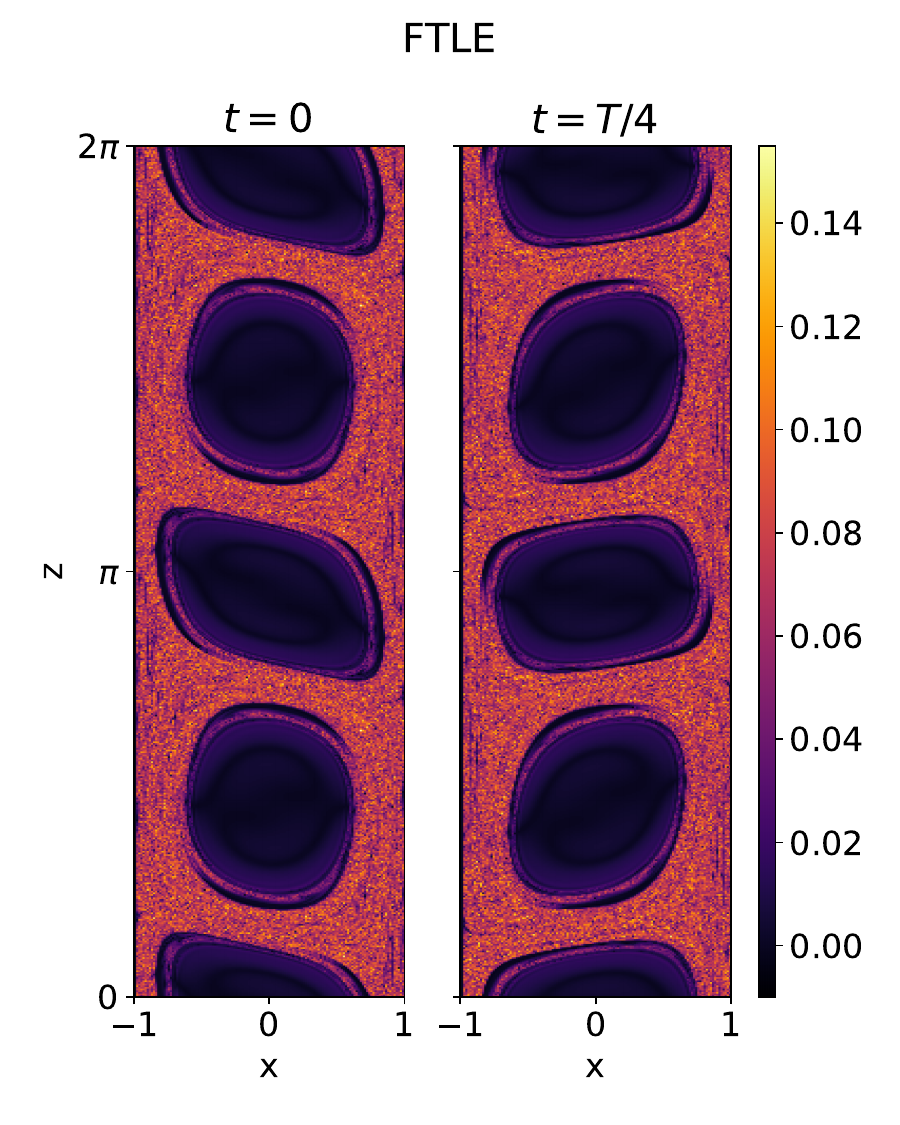}
    \caption{We plot the FTLE as a function of position for $t=0$ and $t=T/4$.
    These results were obtained using particle tracking initialized with the flow states shown in Figure~\ref{big}. 
    Positive FTLEs with sensitive spatial dependence indicate large regions of Lagrangian chaos enveloping the Taylor vortices.}
    \label{ftle}
\end{figure}

The defining criterion of a fast dynamo is that the maximum magnetic growth rate remains positive and finite in the limit $\rm{Rm}\to\infty$.
In Figure~\ref{gamma_rm}, we plot the maximum growth rate as a function of $\rm{Rm}$ for $k_y=0.18$ (shown in green), $k_y=0.29$ (shown in pink), and $k_y=1.0$ (shown in blue).
The $k_y=0.29$ modes are the most unstable at low $\rm{Rm}$ whereas the $k_y=1.0$ modes are  the most unstable for $\rm{Rm}\geq 3.2\cdot 10^5$.
 For all three wavenumbers, we observe no appreciable change in the growth rate for $3.2\cdot 10^5 < \rm{Rm} < 3.2\cdot 10^6$.
The $k_y=0.18$ and $k_y=0.29$ growth rates vary smoothly with respect to $\rm{Rm}$, indicating a single pair of dominant  Floquet modes  are continuously tracked over all $\rm{Rm}$.

These cases provide strong evidence for fast dynamo action, because their growth rates increase smoothly at monotonically decaying rates over five orders of magnitude in $\rm{Rm}$.
For $k_y=1.0$, we observe two cusp points (at approximately $\rm{Rm}=10^3$ and $\rm{Rm}=1.5\cdot 10^4$) where the maximum growth rate's derivative with respect to $\rm{Rm}$ is discontinuous.
These cusp points indicate transitions where multiple distinct  Floquet modes become dominant or sub-dominant as $\rm{Rm}$ changes.

In addition to organizing on a spatial subharmonic with a wavelength twice that of the underlying vortical flow, the dynamo also exhibits temporal subharmonic structure. 
Let $\alpha$ denote the real amplification factor and $n$ the subharmonic index, such that $\mathbf{b}(x,y,z,t+nT) = \alpha^n\,\mathbf{b}(x,y,z,t)$.
Cases with $n=1$ repeat every period $T$, whereas $n=2$ corresponds to a subharmonic oscillation with period $2T$.
We categorize $n$ in Figure~\ref{gamma_rm}, using triangles for $n=1$, circles for $n=2$, and stars for cases with longer periods or quasiperiodicity.
Every unstable case that gives rise to a dynamo has $n=1$ or $n=2$.
For $k_y=0.18$ and $k_y=0.29$, all dynamos have $n=2$, consistent with these two wavenumbers continuously tracking distinct Floquet modes. 
The $k_y=1.0$ dynamos have $n=1$ when $1.5 \cdot 10^3 \leq \mathrm{Rm} < 1.5 \cdot 10^4$.
We observe a transition at the cusp point, where cases with $\mathrm{Rm} \geq 1.5 \cdot 10^4$ have $n=2$.
For $\mathrm{Rm}<1.5 \cdot 10^3$ ($\mathrm{Rm}<100$), the $k_y=1.0$ ($k_y=0.18$) modes decay, including one case with period $3T$ and others with no apparent period.

The exponential stretching of material lines embedded in a flow is a necessary condition for a fast dynamo \cite{vishik_magnetic_1989}.
Lagrangian statistics, such as the Finite-Time Lyapunov Exponent (FTLE), allow us to illustrate the flow's stretching at any point in space and time.
In Figure~\ref{ftle}, we plot the FTLE at $t=0$ and $t=T/4$ using the particle tracking code of \cite{skene_floquet_2023} and algorithm described in \cite{proctor_lectures_1994}.
We use a time horizon of $\Delta t = 200$ with 128 (512) evenly spaced particles in $x$ ($z$).
In large regions of the domain surrounding the boundary layers and separatrices, we obtain positive FTLEs with sensitive spatial dependence.
Comparing this result with the corresponding streamwise velocities (shown in panes A and D of Figure~\ref{big}) reveals that regions of Lagrangian chaos envelop the Taylor vortices.
Lagrangian chaos indicates stretching motions which amplify the magnetic field (shown in panes B, C, E, and F of Figure~\ref{big}).

\begin{figure}[t]
    \centering
    \includegraphics[width=3.4in]{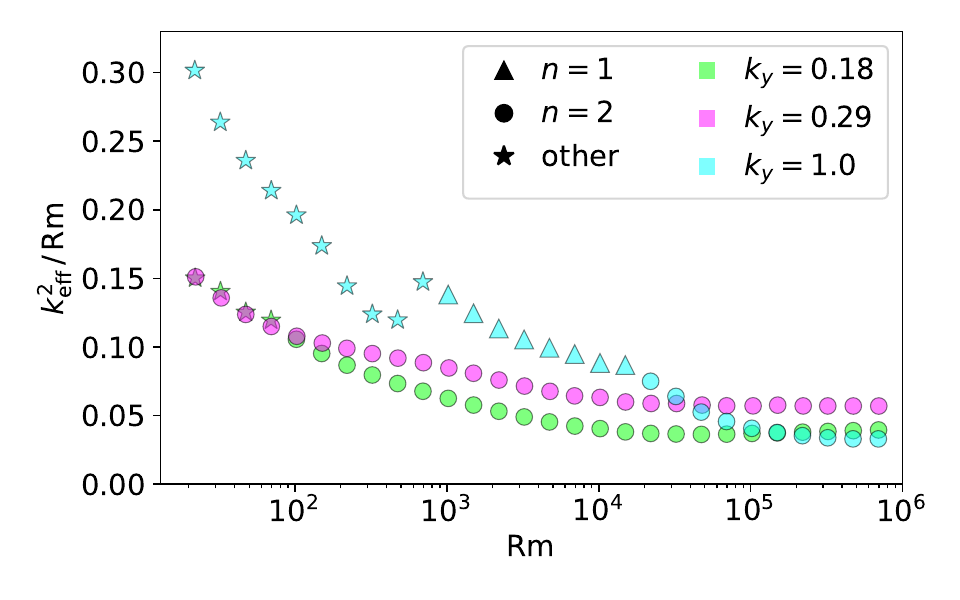}
    \caption{$k_{\rm{eff}}^2 / \rm{Rm}$ as a function of $\rm{Rm}$ for  $k_y = 0.18, \, 0.29,$ and $1.0$.
    This quantity becomes constant for large $\rm{Rm}$, implying $k_{\rm{eff}} \sim \rm{Rm}^{1/2}$.
    Marker and color conventions match those described in the caption of Figure~\ref{gamma_rm}.}
    \label{keff}
\end{figure}

As in \cite{khalzov_fast_2013}, we compute the magnetic field's effective wavenumber $k_{\rm{eff}}=\left( \langle |\grad\times\mathbf{b}|^2 \rangle / \langle |\mathbf{b}|^2 \rangle \right)^{1/2}$ where $\langle \cdot \rangle$ denotes a spatial integral over the problem domain. 
The magnetic Taylor microscale, defined as $k_{\rm{eff}}^{-1}$, measures the magnetic field's typical lengthscale.
In Figure~\ref{keff}, we plot $k_{\rm{eff}}^2 / \rm{Rm}$ as a function of $\rm{Rm}$ for $k_y = 0.18, \, 0.29,$ and $1.0$.
In agreement with Figure~\ref{gamma_rm}, 
the $k_y=0.18$ and $k_y=0.29$ results vary smoothly with $\rm{Rm}$, suggesting a single pair of Floquet modes remain dominant.
For $k_y=1.0$, we observe two non-smooth points, including a discontinuity at approximately $\rm{Rm}=700$ and a cusp point near $\rm{Rm}=1.5\cdot 10^4$, indicating transitions between dominant and sub-dominant modes.
In both cases, $k_{\rm{eff}}^2 / \rm{Rm}$ becomes constant at high $\rm{Rm}$, implying $k_{\rm{eff}} \sim \rm{Rm}^{1/2}$.
The same trend has been observed in slow \cite{perkins_high_1987} and fast dynamos \cite{khalzov_fast_2013}.
Because we observed Lagrangian chaos (Figure~\ref{ftle}) and magnetic structures (Figure~\ref{big}) in large regions surrounding the boundary layers and separatrices, our case resembles the fast dynamo of \cite{khalzov_fast_2013}.

In conclusion, we have demonstrated fast dynamo action sustained by unsteady Taylor--vortex flow. 
Unlike previous studies requiring oscillatory forcing or time-dependent boundary conditions, this fast dynamo arises self-consistently from a time-periodic solution of the Navier--Stokes equations with simple, steady boundary conditions. 
Although experimental realization remains challenging—since the onset of the dynamo investigated in this work occurs when $\mathrm{Rm} \approx \mathrm{Re}$—promising avenues include:
\begin{enumerate}[nosep]
    \item Plasma devices such as \cite{collins_taylor-couette_2014}, which operate near $\mathrm{Pm} \approx 1>\mathrm{Pm}_c\!=\!0.246$ and naturally sustain Taylor--vortex flows with $\mathrm{Re} \approx 150$;
    \item Liquid-metal experiments, where $\mathrm{Pm} \ll 1$ but moderate $\mathrm{Rm}$ can be achieved at large $\mathrm{Re}$.
\end{enumerate}
Both approaches were previously proposed by \cite{gissinger_taylor-vortex_2014}, who also found that adding baffles in the liquid-metal configuration reduces the critical magnetic Reynolds number by 40\% compared with the laminar case. 
The feasibility of liquid-metal experiments is further motivated by \cite{eckhardt_exact_2020}, who showed that turbulent Taylor--Couette flow retains the mean structure and transport properties of the laminar axisymmetric Taylor--vortex state.

Consistent with \cite{gissinger_taylor-vortex_2014}, we observe a spatial subharmonic in the magnetic field whose wavelength is twice that of the underlying Taylor--vortex flow. 
In addition, a previously unreported period-doubling in time appears across three distinct streamwise wavenumbers, $k_y$, showing that this doubled structure is a robust property of the fast dynamo. 
This mechanism provides a minimal route to spatiotemporal scale separation and may represent a key ingredient of large-scale dynamo theory, offering a natural explanation for magnetic reversals, cyclic behavior, and saturation in both laboratory and astrophysical systems.

The authors thank Calum Skene, Steve Tobias, Emma Kaufman, Benjamin Hyatt, and Adrian Fraser for their contributions.
Computations for this work were conducted on the Anvil supercomputer with support by the Rosen Center for Advanced Computing (RCAC) at Purdue University through allocation mth240048 from the Advanced Cyberinfrastructure Coordination Ecosystem: Services \& Support (ACCESS) program, which is supported by U.S. National Science Foundation grants 2138259, 2138286, 2138307, 2137603, and 2138296.
Additional computations were conducted with support by the NASA High End Computing Program through the NASA Advanced Supercomputing (NAS) Division at Ames Research Center on Pleiades with allocation GID s2276.
L.O. is supported by the National Science Foundation Graduate Research Fellowship under grant DGE-2234667.
L.O., D.L., G.M.V., K.C.A., K.J.B., J.S.O., and B.P.B. are supported by NASA HTMS grant 80NSSC20K1280.
D.L., K.J.B., G.M.V., J.S.O., and B.P.B. are supported in part by NASA OSTFL grant 80NSSC22K1738.
D. L. is supported by Sloan Foundation grant FG-2024-21548.
D. L. and F.D. are supported by Simons Foundation grant SFI-MPS-T-MPS-00007353.
E.H.A. is supported by National Science Foundation grant PHY-2309135 and Simons Foundation grant (216179, LB) to the Kavli Institute for Theoretical Physics (KITP).

The code supporting this work is available at \url{https://github.com/liamoconnor9/tvd-fast}.
\bibliography{bib.bib}
\end{document}